\documentclass{elsart}
\usepackage{epsfig}
\usepackage{graphicx}
\usepackage[mathscr]{eucal}

\def\eeq{\end{equation}}
\def\beq{\begin{equation}}
\def\bea{\begin{eqnarray}}
\def\eea{\end{eqnarray}}
\begin{document}
\begin{frontmatter}

\title{Non-extensivity Effects and the Highest Energy Cosmic Ray Affair}

\author{Luis A. Anchordoqui$^a$
and Diego F. Torres$^b$}
\address{$^a$Department of Physics, Northeastern University, Boston,
Massachusetts 02115\\ $^b$Instituto Argentino de
Radioastronom\'{\i}a, C.C.5, 1894 Villa Elisa, Buenos Aires,
Argentina}
\date{\today }
\maketitle

\begin{abstract}
Recent measurements of the cosmic microwave background confirm that
it is described by a Planckian distribution with high precision. It is 
non-extensivity bounded to be less than some parts in
$10^5$, or to some parts in $10^4$ at most. This deviation may
appear minuscule, but may have a non-negligible effect on a 
particle propagating through this background over the course of
millions of years. In this paper we analyze the possible influence of such
a slight deviation upon the propagation of nuclei and protons of
ultra-high energy. These particles interact via photopion and
photodisintegration processes which we examine taking into
account a slight non-extensive background. We show that such a
deviation does not exhibit a significant difference in the energy
attenuation length of extremely high energy cosmic rays.

\noindent {\it PACS number(s): 0.5.20.-y, 98.70.Sa}\\ \noindent

\end{abstract}
\begin{keyword}
non-extensive statistics, cosmic rays
\end{keyword}
\end{frontmatter}
\newpage

The influence of non-extensivity in cosmology,
astrophysics, and cosmic ray physics has been a subject 
of recent interest (see,
for instance, Refs. \cite{attention}). In particular, the effect
of a possible and very slight non-extensivity upon the cosmic
microwave background (CMB) has been thoroughly investigated (see
Refs. \cite{Tsallis3,P2V,RODITI,TORRES-PHYSA98} among others). The
CMB has proven to be described by a Planckian distribution, with a
precision ranging from $\sim 5$ parts in $10^4$ \cite{Tsallis3} to
$\sim 4$ parts in $10^5$ \cite{P2V}. These bounds are given in the
form of constraints in the parameter space, here represented 
by $|q-1|$, where $q$ is the non-extensivity parameter that
appear in Tsallis' extension of statistical mechanics
\cite{TSALLIS-papers}. Within this framework, the generalized photon energy
distribution of Planck's radiation law is \cite{Tsallis3}
\bea \label{PLANCK} D_q(\nu) & = &
\frac{8\pi k^3 T^3}{c^3h^2} \frac {x^3}{e^x-1} (1-e^{-x})^{q-1}
 \;\;\;\;\;\;\;\;\;\;\;\;\;\;\nonumber
\\ & \times &  \left[ 1+ (1-q) x \left(
\frac{1+e^x}{1-e^{-x}}-\frac{x}2
\frac{1+3e^x}{(1-e^{-x})^2}\right)\right], \eea where $x=h\nu/kT$.
Therefore, the expression for $D_q(\nu) c^3h^2/8\pi k^3 T^3$
generalizes the Planck's law, and in the limit of $x\ll 1$, the
Rayleigh-Jeans' law. For $h \nu |1-q| \gg kT$, the $(1-q)$ correction
diverges, which renders the  expansion invalid. For $q \sim
1$, because of
the suppression of the exponential factor,  
the divergent region has no practical importance. Consequently, the total
emitted power per unit surface is given by \beq P_q \propto
\int_0^\infty D_q d\nu \sim  \sigma_q T^4,\eeq where a
$q$-dependent Stefan-Boltzmann constant has appeared.
The  thermal spectrum given in Eq. (\ref{PLANCK}) was
applied to the CMB in order to test for deviations from Planck's
law \cite{Tsallis3,P2V,RODITI,TORRES-PHYSA98}.
However, the influence of such a possible deviation has not been
yet fully explored. In this letter, we aim to look for any possible
non-extensivity effects in the propagation of ultra-high
energy cosmic rays (CRs).

CRs do not travel unhindered through intergalactic space, as there
are several processes that can degrade the particles' energy. In
particular, the thermal photon background becomes highly
blueshifted for ultra-relativistic protons. The reaction sequence
$p\gamma \rightarrow \Delta^+ \rightarrow \pi^0 p$ effectively
degrades the primary proton energy. This provides a strong constraint on
the proximity of CR-sources, a phenomenon known as the
Greisen-Zatsepin-Kuz'min (GZK) cutoff \cite{gzk}. Specifically, less
than 20\% of $3 \times 10^{20}$ eV ($1 \times 10^{20}$ eV) protons
can survive a trip of 18 (60) Mpc. For nuclei and photons the
situation is, in general, more drastic \cite{puget}.\footnote{It
should be stressed that super-heavy nuclei can break the GZK
barrier \cite{a}. However, the abundance of nuclei heavier than
iron (in the cosmic radiation) is expected to be smaller by 3 or 5
orders of magnitude relative to the lighter ones.} In recent
years, a handful of extensive air showers have been observed with
nominal energies at or above $10^{20} \pm 30\%$ eV \cite{yd}. The
arrival directions of the primary particles are distributed widely
over the sky, with no plausible astrophysical sources within the
GZK distance ($\approx 50$ Mpc) \cite{hillas}. Furthermore, although 
the first five
events with $E> 8 \times 10^{19}$ eV (at 1-standard deviation) did
in fact point toward high redshift compact radio-loud quasars
(astrophysical environments that could accelerate CRs above the
GZK energies via shock mechanisms) \cite{FB}, with the inclusion
of subsequent data this association is controversial
\cite{sigletal}. Certainly the energy window of ``super-GZK'' CRs
greatly exceeds that of man-made accelerators, and so speculation
on new high-energy physics is open \cite{BS}.

The interaction length of a nucleon to undergo photopion production
is $\approx 6$ Mpc, whereas a typical nucleus 
suffers disintegration (against photons of the microwave and
infrared radiation backgrounds), losing about 3-4
nucleons per traveled Mpc \cite{puget}. Consequently, if taking $q \neq 1$
leads to unobservable effects in the nucleus photodisintegration
rate, this also implies unobservable effects in the photopion
production process. For iron nuclei with a Lorentz factor $\log
\Gamma \geq 9.3$, the CMB provides the target photon field, hence
they are excellent candidates to test whether the inclusion of 
non-extensivity effects may alter the ``average particle kinematics''.
In the universal
rest frame (in which the microwave background
radiation is at $2.73$ K), the disintegration rate $R$, of an
extremely high energy nucleus of iron ( $^{56}$Fe),
propagating through an isotropic
soft photon background of density $n$, is given by \cite{S},
\begin{equation}
R = \frac{1}{2\Gamma^2} \int_0^\infty d\epsilon\,
\frac{n(\epsilon)} {\epsilon^2}\,\int_0^{2\Gamma\epsilon}
d\epsilon' \,\epsilon' \, \sigma (\epsilon'),
\end{equation}
where $\sigma$ stands for the total photo-disintegration cross
section. According to the photon energy in the nucleus rest frame
($\epsilon'$) one can distinguish different regimes. From
$\epsilon' \approx 2$ MeV up to $\epsilon' \sim 30$ MeV, the total
photon absorption cross section is characterized by a broad
maximum, designated as the giant resonance, in which a collective
nuclear mode is excited with the subsequent emission of one (or
possible two) nucleons. Beyond the giant resonance and up to the
photopion production threshold, i.e. $\epsilon' \approx 150$ MeV,
the excited nucleus decays dominantly by two nucleon
(quasi-deuteron effect) and multi-nucleon emission. The giant
dipole resonance can be well represented by a Lorentzian curve of the form
\cite{kt},
\begin{equation}
\sigma(\epsilon') = \sigma_0 \frac{\epsilon'^2 \,
\Gamma_0^2}{(\epsilon^2_0 - \epsilon'^2)^2 \, + \,\epsilon'^2 \,
\Gamma_0^2},
\end{equation}
where $\sigma_0 = 1.45 A$ mb, and $\Gamma_0 = 8$ MeV is the energy
bandwith. The peak energy depends on the mass number of the
nucleus: $\epsilon_0 = 42.65 A^{-0.21}$ MeV for $A>4$, and
$\epsilon_0 = 0.925 A^{2.433}$ MeV for $A\leq 4$. Above 30 MeV the
cross section is roughly a constant, $\sigma = A/8$ mb.
In Fig. 1 we show the variation of $dR/d\epsilon$ as a function of the
background photon energy for different values of the
non-extensivity parameter in $n(\epsilon)\equiv D_q(\nu)/hv$.
It is important to stress that $1-q = 10^{-5}$ is admitted for
the most restrictive
cosmological bounds existing today: those obtained from the CMB.
The effects of $q \neq 1$ are a small decrease in the peak value
and a shorter tail.

At this stage, it is noteworthy  that the
nucleus emission is isotropic in the rest system of the nucleus.
Moreover, for photodisintegration the Lorentz factor is conserved
during propagation.\footnote{Note that processes which involve the
creation of new particles which carry off energy (pair production,
photopion production) do not conserve $\Gamma$.} Thus, the
photodisintegration average loss rate reads, \beq \frac{1}{E}
\frac{dE}{dt} = \frac{1}{A} \frac{dA}{dt}, \eeq
and the total energy loss time $\tau$ is given by
\begin{equation}
\tau^{-1} \equiv \frac{1}{E} \frac{dE}{dt} = \frac{1}{\Gamma}
\frac{d\Gamma}{dt} + \frac{R}{A}.
\end{equation}
Above $2 \times 10^{20}$ eV the reduction in $\Gamma$ (due to
$e^+e^-$ production) can be safely neglected \cite{puget}, so
using the scaling for the cross section proposed in \cite{puget},
\beq \left. \frac{dA}{dt}\right|_A \approx \left.
\frac{dA}{dt}\right|_{{\rm Fe}} \frac{A}{56} =
-\frac{R(\Gamma)|_{{\rm Fe}}\, A}{56}, \eeq it is
straightforward to obtain the photodisintegration history in terms
of the nucleus time flight: \beq A(t) = 56 \exp[ - R|_{\rm Fe}
\,\,t /56]. \eeq
It is easily seen that although non-extensivity effects are observable in the
derivative of $R$, they disappear when one integrates over the
energy. Even if we artificially enhance the difference in the
value of $|q-1|$, and take for instance $q=0.95$ (see Fig. 2), the
mean value of the nucleon loss is within the errors of a
Monte Carlo simulation \cite{phd}.\footnote{The case could be made for such 
high values of $|q-1|$ in the early universe (before
the time of primordial nucleosynthesis) but they are certainly not in
agreement with CMB measurements.}

In summary, photopion and photodisintegration in a non-extensivity framework do
not differ from the standard case (when non-extensivity is bounded
as explained) even when nuclei or protons are compelled to
interact with the background through a significant part of the
size of the entire universe.

In closing, we would like to point out that the most efficient
target for extremely high energy gamma rays to produce $e^+e^-$
pairs are background photons of energy $\sim m_e^2/E \leq 10^{6}$
eV $\approx 100$ MHz (radio-waves), with a mean free path of 20 -
40 Mpc at $3 \times 10^{20}$ eV (see Protheroe-Johnson in Ref.
\cite{puget}). However, at $10^{15}$ eV the energy loss rate is
dominated by collisions with the CMB. The relic photons drop the
energy attenuation length down to $10^{-2}$ Mpc. Then, from the
phenomenological point of view, it would perhaps be interesting to
analyze whether non-extensivity effects yield any observable
deviation on the energy loss time of photon-photon interaction.

\section*{Acknowledgments}

We would like to thank Tom McCauley for a critical reading of 
the manuscript.
Remarks by an anonymous referee are gratefully acknowledged.  
This work was supported by CONICET, and Fundaci\'on Antorchas 
through a grant to D.F.T.

\newpage

\begin{figure}[t]
\label{hp}
\begin{center}
\epsfig{file=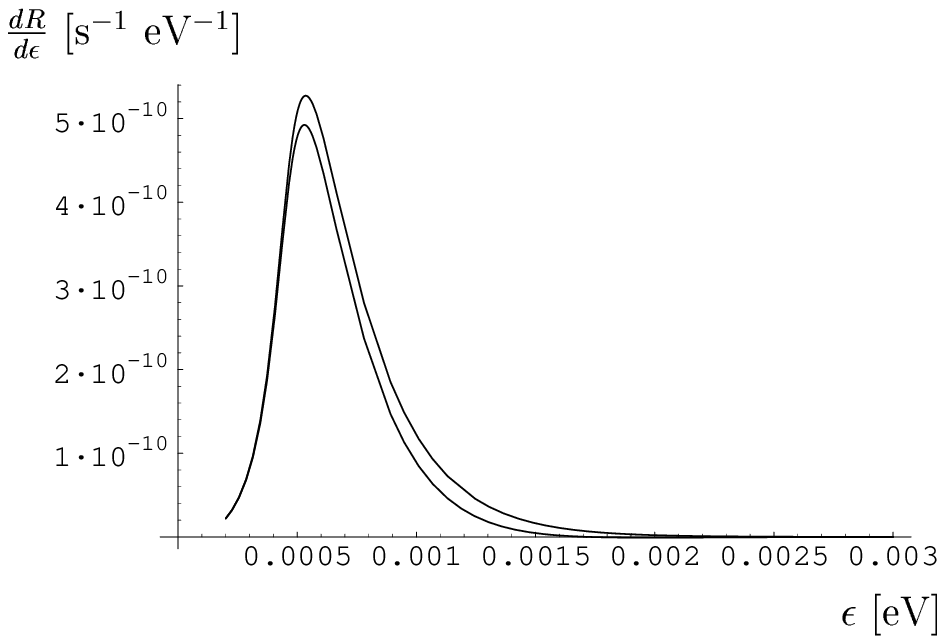,width=9.cm,height=9.cm,clip=}
\caption{Variation of $dR/d\epsilon$ with respect to the background photon
energy for different values of $q$ and $\log \Gamma = 10.3$.
The curve with the higher peak stands for the typical Planckian spectrum,
whereas the other curve owns a deviation from Planck spectrum
of $1-q \equiv 10^{-5}$.}
\end{center}
\end{figure}

\begin{figure}[t]
\label{hp2}
\begin{center}
\epsfig{file=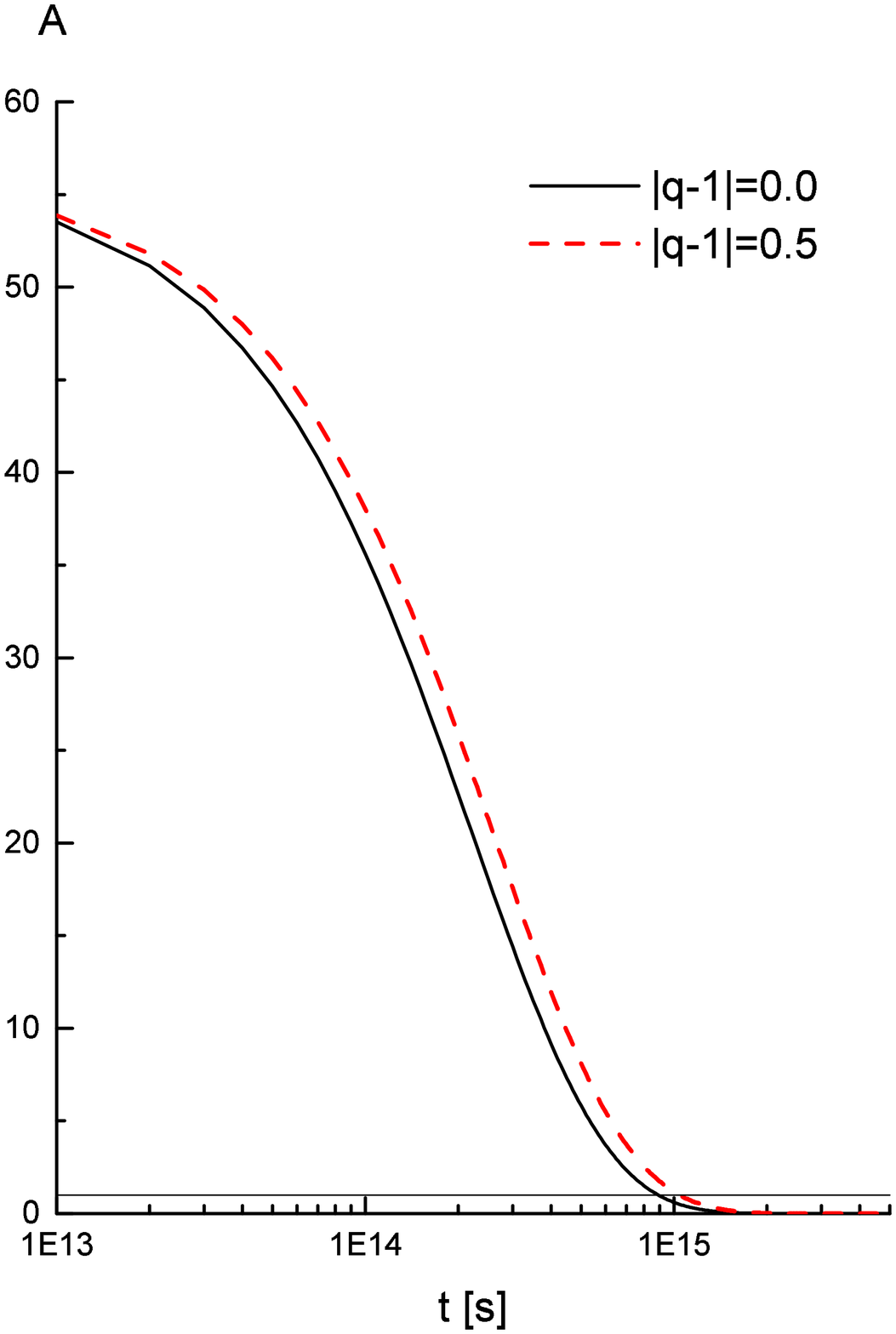,width=9.cm,height=9cm,clip=} \caption{
Evolution of the mass number $A$ of the surviving fragment vs.
travel time for injection energy $\log \Gamma = 10.3$. It is
important to keep in mind that 1 Mpc $\equiv$ $1.03 \times
10^{14}$ s $\equiv$ $3.26 \times 10^{6}$ light years $\approx 3
\times 10^{19}$ km.}
\end{center}
\end{figure}

\end{document}